# Convolutional neural network for detection and classification of seizures in clinical data

Tomas Iešmantas[1], Robertas Alzbutas
Kaunas University of Technology
Department of Mathematics and Natural Sciences

## Abstract

Epileptic seizure detection and classification in clinical electroencephalogram data still is a challenge, and only low sensitivity with a high rate of false positives has been achieved with commercially available seizure detection tools, which usually are patient non-specific. Epilepsy patients suffer from severe detrimental effects like physical injury or depression due to unpredictable seizures. However, even in hospitals due to the high rate of false positives the seizure alert systems are of poor help for patients as tools of seizure detection are mostly trained on unrealistically clean data, containing little noise and obtained under controlled laboratory conditions, where patient groups are homogeneous, e.g. in terms of age or type of seizures.
In this study authors present the approach for detection and classification of a seizure using clinical data of electroencephalograms and a convolutional neural network trained on features of brain synchronisation and power spectrum. Various deep learning methods were applied, and the network was trained on very heterogeneous clinical electroencephalogram dataset. In total, eight different types of seizures were considered, and the patients were of various ages, health conditions and they were observed under clinical conditions. Despite this, classifier presented in this paper achieved sensitivity and specificity equal to 0.68 and 0.67, accordingly, which is a significant improvement as compared to the known results for clinical data.

## Introduction

Almost two decades ago, it was observed that 63% of patients under medication in an epilepsy service were rendered as seizure-free patients [45]. Since then little has changed, despite new antiepileptic drugs became available in the market, it was found [46] that the percentage of patients getting rid of seizures have changed only slightly. Better drug delivery systems to the patient organism might help to increase the efficiency of epilepsy treatment. In addition, considering the possible toxicity of the epilepsy drugs [25], a smart system, able to predict seizure in advance and deliver a small dosage of drugs to interrupt the development of that seizure, would be a significant addition to the arsenal. At least one could expect a system, able to detect the onset of the seizure and inform patient or medical personnel about it – this would help to reduce the stress or development of more severe physical or mental conditions like depression [26].
In fact, many such systems have already been patented. Patents describe various devices, able to detect and predict the developing seizures using nonlinear chaotic time series techniques [27], or by tracking the time evolution of chaotic entrainment [28]. Another concept uses a wavelet transform maximum modulus algorithm applied to data from the entire body [29] or tracks impedance of the brain [30]. These examples barely scratch the surface of many patented tools (devices/methods/algorithms) able to detect (at least for the dataset they were validated) and predict seizures.

[1] Corresponding author: tomas.iesmantas@gmail.com

A large number of patents correlates with an even larger set of research reporting high accuracy of seizure detection and prediction. For example, F. Morman et al. [31] reported results of seizure detection for five epilepsy patients as follows: Area Under the Curve (AUC), as a measure of the accuracy of the algorithm, equal to 0.83, with sensitivity and specificity measure correspondingly equal to 0.95 and 0.65. This algorithm was trained on intracranial EEG data first extracting so-called mean phase coherence, more commonly known as phase locking value, which measures the degree of brain synchronisation. M. J. Cook et al. [32] trained a classification algorithm on intracranial EEG data of 15 patients and obtained sensitivity 0.78, although authors did not report what kind of EEG dataset features were used. M. D'Alessandro et al. [33] considered prediction of seizures for four patients based on intracranial EEG data – authors reported sensitivity equal to 0.625 and specificity equal to 0.91; numerous features were calculated every 1 minute with 2.5 second displacement and extracted considering the time domain, frequency domain, and nonlinear dynamics.

Often used Freiburg dataset consisting of 21 patient (data collected at Epilepsy Center of the University Hospital of Freiburg, Germany and no longer freely available, http://epilepsy.uni-freiburg.de/freiburg-seizure-prediction-project/eeg-database ) was used in various studies with variable, although very high accuracies:

- [34] reports 0.975 sensitivity of algorithm, trained on spectral power features (20 second windows of intracranial EEG).
- [35] obtained 0.71 sensitivity with perfect specificity. Authors used convolutional neural networks on various features (20 minute window).
- [37] reports 0.9 sensitivity and 0.21 false positives per hour.

High accuracy was reported on scalp EEG as well by A. Shoeb [36], where the author considered seizure prediction for 22 pediatric patients and sensitivity equal to 0.9 was obtained. On the same dataset (from scalp EEG), S. Nasehi and H. Pourghassem [37] obtained sensitivity equal to 0.98 with 0.06 false positives per hour.

The research on seizure detection is very active and during recent years there have been many different approaches. Application of convolutional neural networks is being applied to this problem quite extensively nowadays with very promising results (see Table 1).

*Table 1 Convolutional neural network based seizure detection from EEG features*

| Reference (year) | Accuracy measures (Accuracy – A, Sensitivity – Se, Specificity - Sp) | Time window length, seconds | EEG type |
|---|---|---|---|
| [53] (2018) | A=88.67%, Se=95%, Sp=90% | 23.6 seconds | Intracranial (Freiburg) |
| [54] (2019) | Se=86.6%, 0.84 false alarms per hour | 5 seconds | Intracranial (CHB -MIT) |
| [55] (2019) | Se=89.4%, Sp=97%, 0.6 false alarms per hour | 30 seconds | Scalp |
| [56] (2018) | Se=88.9%, Sp=93.78% | 10 seconds | Scalp (laboratory conditions) |
| [57] (2019) | Area Under Curve = 0.9 (Se and Sp not reported) | 1 second | Intracranial (Mayo Clinic) |

From this brief review of seizure detection results, one can make several observations: most of the results are reported for intracranial EEG data, records are obtained under laboratory conditions, often during long-term monitoring for presurgical evaluation. High accuracies were achieved for seizure vs non-seizure classification algorithms for intracranial EEG. Features, representing a state of the brain, are often calculated using time windows significantly larger than 1 second, even though this is a time resolution often used in the fine-grained analysis of EEG [38].

EEGs are not the only option to use in seizure reporting. Other modalities may also be used for this purpose, like accelerometry [48], extracerebral movements [49], audio recordings [50], worn wrist devices [52], etc. (more broad review of the methods can be found in [51]). However, in this paper, the focus is on EEG signals.

However, given all the advances in seizure prediction and all the patents for seizure detection/prediction systems, then one may ask why commercially available tools perform only at around 0.3 sensitivity with high false positives rate [39]. The answer probably lies in the fact that most of the results of various classifiers were obtained for "nice" EEG data, i.e. EEG obtained under laboratory conditions when seizures observed are often of the same type, and they are within homogeneous age groups. A similar concern was recently voiced by D. R. Freestone *et al*. [40], where authors noted that "data that are recorded in the epilepsy monitoring unit are not representative of 'normal' seizures". Also, as observed above, many algorithms were evaluated on intracranial EEG data, and it is known that intracranial electrodes can change the dynamics of seizures [41].

A recently released dataset and still actively under further development as Temple University Hospital (TUH) EEG Corpus [9], which is the world's largest publicly available database of clinical EEG data, might fill the void of more realistic seizure data – it was obtained under clinical conditions, spans over different age groups, different types of seizure events, different health conditions of patients. The first reported benchmark based on this database produced the following results [38]: sensitivity – 0.292, specificity – 0.667. Authors used a classification algorithm based on the hidden Markov chain model. The low accuracy of these methods shows their inadequacy for this TUH EEG Corpus. However, it is the first baseline to overcome for other machine learning algorithms.

Hence, the purpose of this study was to probe this clinical reference database with algorithms, which have shown high accuracy of seizure detection on other specific datasets. More specifically, we selected convolutional neural network as a method, that showed most promise when applied to other datasets (see Table 1 above). We adapted the pattern construction from EEGs in order to be able to use very short 1 second epochs, as opposed to most of the other studies. The main contribution of this paper is application and investigation of convolutional neural network architectures to the very large and highly heterogeneous EEG dataset. It must be stressed, that this dataset was obtained not under laboratory conditions, but in various intensive care units. Proposed approach of classification enabled to achieve sensitivity and specificity equal to 0.68 and 0.67, respectively – a considerable improvement from the known result on this dataset.

The structure of the paper is as follows: in the section for material and methods, the considered TUH EEG Corpus database is described in details. Also, features of brain synchronisation, entropy, and power spectrum, as well as brain activity patterns and methods relevant for the considered convolutional neural network is defined and described. In the section for results and discussion, an extensive analysis of TUH EEG Corpus by the convolutional neural network classifier is presented.

# Material and methods
### Data
As was pointed out in the introduction section, most seizure detection and prediction algorithms were validated on the relatively "nice" and homogeneous EEG datasets. However, one clearly cannot expect to have always the same type of seizures and the same conditions under which EEGs are recorded – clinical setting produces much more variable EEG recordings (influenced by environment, equipment, electrode location, noise, etc.) as opposed to the tightly controlled research setting. Therefore, there was a need for a dataset, which would become a testing ground for the machine learning algorithms. The Temple University Hospital EEG Corpus [9] could be such a reference dataset. The entire corpus contains 16,986 sessions from 10,874 unique patients, and each session additionally includes one physician report.

Part of the corpus is devoted specifically for the seizure detection problem (including classification by seizure types). It is also subdivided into two segments: training data and evaluation data. Training dataset is comprised of 196 patients with total seizure time interval being equal to 14 hours and non-seizure time interval – 244 hours. The evaluation dataset is constructed from recordings of 50 unique patients with total seizure and non-seizure time intervals being equal to 15 hours and 152 hours correspondingly. Furthermore, there were several types of seizures observed in the training and evaluation datasets (see Table 2), which also adds up to the variability of the dataset.

*Table 2 Types of seizures considered in the training and evaluation of TUH EEG Corpus dataset*

| Type of seizure | Brief description |
| --- | --- |
| 1. Focal Non-Specific Seizure | Focal seizures which cannot be specified by its type |
| 2. Generalised Non-Specific Seizure | Generalised seizures which cannot be further classified into one of the groups below |
| 3. Simple Partial Seizure | Partial seizures during consciousness; Type specified by clinical signs only |
| 4. Complex Partial Seizure | Partial seizures during unconsciousness; Type specified by clinical signals only |
| 5. Absence Seizure | Absence discharges observed on EEG; patient loses consciousness for a few seconds (Petit Mal) |
| 6. Tonic Seizure | Stiffening of the body during a seizure (EEG effects disappears) |
| 7. Clonic Seizure | Jerking/shivering of the body during a seizure |
| 8. Tonic-Clonic Seizure | At first stiffening and then jerking of the body (Grand Mal) |

\* See webpage https://www.isip.piconepress.com/projects/tuh_eeg/downloads/tuh_eeg_seizure/v1.1.1/

The data set is highly imbalances: only 6 % of all the data points are annotated as belonging to one of the seizures described in Table 2. This affects sensitivity as well as specificity severely. Therefore these measures will be reported together with over all AUC (Area Under Curve) measure, which is not sensitive to the balance of data.

**Seizures: synchronisation, entropy and power spectrum**
In this subsection, several features of EEG dataset are defined. In previous studies, these features demonstrated as carrying a significant amount of information regarding the differentiation between seizure and non-seizure events. Namely, phase locking value [42], average power and entropy for different EEG frequency bands. After the definition follows the initial exploration of these features regarding distributions in the seizure/non-seizure classes.
One of the most investigated aspects of seizures is the phenomena of synchronisation. Spontaneous seizures have been interpreted as abnormal excitability and synchronisation of large neuronal populations [10, 12]. However, recent advances have shown [11] that the picture is more complicated: desynchronisation at the onset of the seizure has been observed in neural activity both in vitro and in vivo [13, 14]. This initial desynchronization is followed by the increase of the synchronisation until it reaches its peak during the final moments of the seizure and at last the seizure simultaneously terminates over large areas of the brain [15, 16]. Thus, there is a complex interplay of synchronisation and desynchronization of the brain during the evolution of the seizure.
On the other hand, the seizure is not the only time when synchronisation occurs. In general, it has been hypothesised [17], that synchronisation phenomena are inherent in the brains large-scale integration. For example, in the visual binding problem [18], when the different attributes of an object brought together in a unified representation given that its various attributes – edges, colour, motion, texture, depth and so on – are treated separately in specific visual areas [18]. Also, the increase in synchronisation has been observed during the following activities: flicker stimulation [19], process related to pattern vision [20] and reading [21]. Given this brief review, it could be hypothesised that synchronisation features possibly are not very strong predictors as in the cases with more homogeneous seizure EEG datasets. Also, the relative success of such features, as described in the introduction, probably is simply due to the aforementioned EEG data cleanliness, which is not the case in TUH EEG Corpus case.

One of the commonly used measures of dynamic system synchronisation is the *phase locking value* (PLV). The locking of phases of two oscillators occurs when the phase entrainment condition is satisfied:

$$\phi = |\varphi_1 - \varphi_2| < const.,$$

where $\varphi$ is the instantaneous phase of the oscillator, and $\phi$ is their relative phase. Phase entrainment condition can be defined more generally, but this definition is sufficient for our purposes. Instantaneous phases can be estimated in various ways, but two methods are mostly used: a signal is transformed by Hilbert transform or using complex Gabor wavelets. However, it was showed [22] that these methods are equivalent. For our calculations, the Hilbert transform method was chosen, i.e. initially the original signal $x(t)$ is filtered by a band pass filter $g$ and then this filtered signal $x_g(t)$ is Hilbert-transformed to obtain a complex signal $X_g(t) = x_g(t) + iH\{x_g(t)\}$. Instantaneous phase is $\varphi = \arg(X_g)$.

Phase locking value is defined as follows

$$PLV = \left| \frac{1}{N} \sum_{n=1}^{N} e^{i\phi_n} \right|, \qquad (1)$$

where $\phi_n$ is the relative phase at an n$^{th}$ time point; $N$ is the length of a time interval over which averaging is performed.

Another measure that quantifies the strength of synchronisation is based on *Shannon entropy* and is defined [23] as

$$\rho = \frac{S_{max} - S}{S_{max}}, \quad S = -\sum_{k=1}^{K} p_k \ln p_k, \qquad (2)$$

where $S$ is the entropy of the distribution of the relative phase $\phi$, $S_{max} = \ln K$, $K$ is the number of bins.

Another kind of features is spectral covariance or coherence. However, these features will not be used in the study. It is due to the fact that these features are suitable only for stationary signals [22], and one cannot expect this stationarity from EEG records.

In addition to the PLV and $\rho$ features, a *power spectrum* of EEG is taken into consideration. Power spectrum was divided into seven frequency bands (i.e. delta, theta, alpha, low beta, high beta, low gamma, and high gamma) and logarithms of averages over each of the band is calculated. The reasoning behind the power spectrum consideration as an informative feature is the changes in spectral power during various phases of the seizure evolution, e.g. it has been observed the increase of spectral power of high frequencies during the start of the seizure [24].

**Patterns of brain activity**
In the introduction, the importance of short detection window (1-2 seconds, [38]) was discussed as most of the algorithms rely on the longer time windows: lengths vary from 20 seconds [34] to 5 minutes [35]. Broader detection horizons (or epochs) enable to extract more features from time subintervals in one epoch. Thus, a single feature vector embodies not only a momentary state of the brain during the seizure but also the specific evolution of it (see the discussion above and in [11] for the different phases of synchronisation changes). On the other hand, 1-second window (as a length of detection horizon is selected in this study) is too short to reflect significant changes of EEG before, during and after the seizure. Hence, authors were forced to use features that do not reflect the time evolution of EEG signal, i.e. each time-dependent element in the vector for any single feature reflects the average state over this short detection horizon. This poses a great challenge - the shorter the period, the smaller the amount of information is contained in it; also, short time periods are the reason for higher variance of feature vectors as obtained using different time windows. A notable example of detection of seizures for very short time windows is the Kaggle challenge [43], where 1-second non-overlapping time windows were also used for seizure detection, and very high accuracies were achieved even with such short time slices of specific EEGs.

Because the combinations of features (as pointed out in the introduction part) play a significant role in the efficiency and accuracy of classification algorithms, in this study, they are used additionally. A feature matrix, rather than feature vector, is extracted from each 1-second time window of EEG –such a matrix is called *a pattern*. The same name is also used by other authors [35] when combinations of features are formed for application of their convolutional neural network (more details follow below). However, authors [35] had the advantage of having very long detection horizons (5 minutes): they divided the entire horizon into smaller windows, and each bivariate feature vector is placed into separate pattern matrix column (i.e. columns represented the time evolution of features). In this paper, the construction of patterns is adapted for short time windows.

Consider 1 second (or any other length) epoch of EEG signal matrix $X \in \mathbb{R}^{n \times t}$, where $n$ is the number of EEG channels and $t$ is the number of discrete time measurements, which in our case is equal to 250 points – 250 Hz was the frequency at which the EEG sampled, at least in the TUH EEG Corpus portion which was devoted to the seizure detection.

Three types of features were calculated for each 1 second epochs (no intersection). For the PLV and Shannon entropy features:
1. first each EEG channel was filtered in 7 frequency intervals ( [44], delta (< 4 Hz), theta (4–7 Hz), alpha (7–13 Hz), low beta (13–15 Hz), high beta (14–30 Hz), low gamma (30–45 Hz) and high gamma (45–70 Hz));
2. then for each filtration result, Hilbert transform $H(\cdot)$ was calculated from which instantaneous phase $\varphi$ was extracted. In other words, from one second of EEG of one channel, one gets 7 times series of instantaneous phases.
3. The pairwise channel differences of instantaneous phases are calculated to get relative phases. This is done for each of 7 filtration intervals. And then PLVs and Shannon entropies are calculated for each result, collapsing a series of 250 points to a single number.

So, for example suppose there are 10 channels. Then for each 1 second time series there will be $7 \cdot 10 = 70$ frequency filtration results. The Hilbert transform is performed and instantaneous phases extracted. Then, take instantaneous phases 10 channels corresponding to, say, delta band filtration and obtain $\frac{10 \cdot 9}{2} = 45$ pairwise differences – these differences will be relative phases. Then for each of 45 relative phases calculate PLV and Shannon entropies. In other words, there will be $45 \cdot 7 = 315$ PLV values and 315 Shannon entropy values.

In order to form a patter, a picture of the brain state during the particular 1 second epoch, feature values are represented not as a vector but rather as symmetric square matrix. For the above example, those 45 values of PLV and Shannon entropy values are put into symmetric $10 \times 10$ format matrix (where entry with address $(i, j)$ is the feature value obtained from relative phase of channels $i$ and $j$).

These patterns are then stacked together on top of each other. Referring to the example, final output for 1 second epoch will be two PLV and Shannon entropy matrices of dimensions $(7 \cdot 10) \times 10 = 70 \times 10$.

To form power spectrum features, take the pairwise differences of each channel, obtain power spectrum for 1 second epoch and then obtain averages of spectrum values over the 7 band intervals. Again, relating to the above example, the final result would be $70 \times 10$ matrix. These matrices are calculated for each second of the EEG for each patients.

The above-constructed patterns were then used by the learning algorithm. Several feature combination settings were considered for training the algorithm: using only one feature, using pairs of features and using all features together. In the setting when two or three features are used then their matrices simple are stacked by sides (e.g. if we combine PLV and Shannon entropy feature matrices of dimensions $70 \times 10$, then we'll get after combining a matric of size $70 \times 20$). More details on the learning and convolutional neural network (CNN) are given in the next subsection. The reasons of more extensive CNN presentation are that other classification methods like Support Vector Machines or k-nearest neighbours algorithms were already applied in many seizure detection cases, while CNN with specific features is a newer concept in this type of problems (although CNN has a long history in image recognition applications [1, 2]). In addition, this classification method for long time window showed a significant superiority over other methods in seizure classification task (see the reference [35]). So, it is important to focus on it and describe it in greater details keeping in mind the importance and issues for short time windows.

**Convolutional neural network**
As it was pointed out in the previous subsection, convolutional neural networks are superior over logistic regression and support vector machines – both methods, often in the bucket of machine learning methods in this context. Thus, it is reasonable to introduce CNN in greater detail to facilitate more widespread applications of it within the EEG dataset classification community.

Historically, CNN was developed for image classification and was based on the theory (which is now outdated) on how visual cortex works. The significant turning point for the theory of CNN was several seminal works by Yann LeCun [1, 2] in which the application of CNN was made for the recognition of handwritten digits. The original image was convolved with several filters impacting the parts of small size (usually it is 2x2, 3x3, 5x5, etc.) to form layers of feature maps. Usually, nonlinear functions like rectified linear unit (ReLU) or sigmoid are applied to the convolved images. Then these feature maps are subsampled (often by a factor of 2) to form the next layer – at first it may look like a reduction of useful information, but in fact, it facilitates the generalization process. One may put as many convolution/subsampling layers as it is necessary. In addition, one can put fully connected layers at the end of the network. A general structure of the convolutional neural network is presented in Figure 1 below.

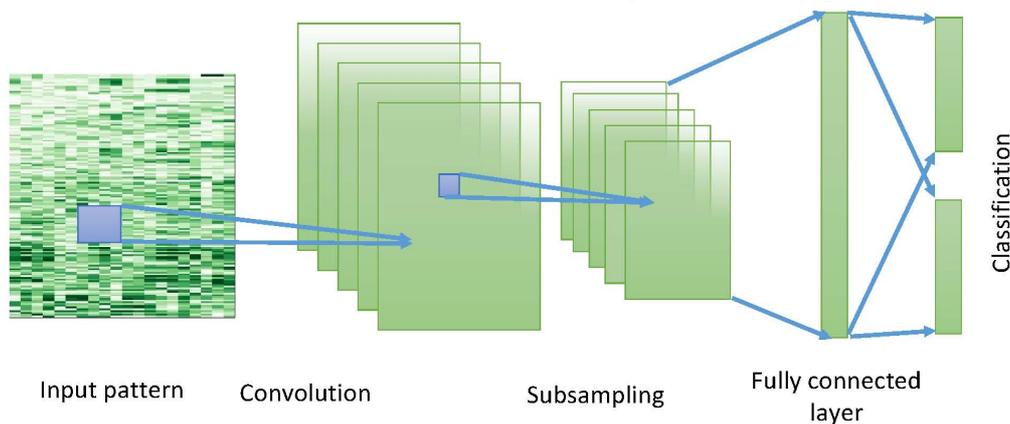

*Figure 1 Visualization of simple architecture of convolutional neural network: 1 convolutional, one subsampling and two fully connected layers, including the final classification or output layer. The convolutional layer is obtained by convolving the image with some number of small filters. Most often the size of filters are 3x3, 5x5, 7x7, i.e. very small as compared to the whole image. Subsampling serves as dimensionality reduction operation with the extra advantage being that this operation makes the network position invariant. The last layer produces a two-class probability for non-seizure image and seizure image.*

Main ingredients of a convolutional neural network are:

1. A filter. This is a small matrix (usually of dimensions $3x3, 5x5, 7x7$), with unknown (therefore, learnable) entries. Filter entries can be view as neurons in a classical neural network, the only thing is that they have shared parameters. The parameter sharing means that rather than learning separate set of parameters for every location, only one set is learned. Most often the numbers of filters in each convolutional layer are 8, 16, 32, etc.
2. Discrete convolution operation is used to convolve the input (initial pattern, if it is the first layer, or output from the previous layer) with filters. An analogous operation in classical (or densely connected) artificial neural networks is the multiplication of input by a weight matrix.
3. Nonlinear activation function is applied after the convolution with filters. In recent CNN application, the so called rectified linear activation function is used. It became popular after it was shown to enable faster network training on a large scale image recognition tasks [58].
4. Pooling operation. In order to decrease the dimensions of a convolutional layer, one uses so called pooling operations. This operation replaces the output of a layer at a certain location with a summary statistics of the neighboring outputs. Most often are used two such summary statistics: maximum of a neighborhood or an average.

There are no restrictions on how many filters should be within one layer and how many convolutional layers should one put. However, this freedom of choice is not a very good thing, because one there are also no guidelines on what architecture will best fit the problem at hand. Most often several architectures significantly differing from each other (e.g. obtained by doubling the number of convolutional layers) are evaluated against each other by 10 fold cross-validation procedure: divide the data set into 10 equal parts, use 9 parts for training and 1 part for training and repeat 10 times – the final architecture is selected by the largest average accuracy.

The training could be done by the variant of the backpropagation algorithm or with gradient-based deep learning environments like TensorFlow [3]. In this study, TensorFlow is used together with Adam optimiser [4] and learning rate equal to 1e-3. To obtain nonlinearity at each convolutional layer the ReLU (Rectified Linear Unit) function $f(x) = \max(0, x)$ is used. Later, for the classification layer, SoftMax function with two classes is used. Then the output from this last layer is provided as input to the cross-entropy loss function. If the output of SoftMax function is a vector reflecting probabilities of K classes $\tilde{y} = (\tilde{y}_1, \tilde{y}_2, ..., \tilde{y}_K)^T$, then the loss due to one pattern is equal to:

$$L(y, \tilde{y}) = -\sum_{j=1}^{K} y_j \log \tilde{y}_j,$$

where $y = (y_1, y_2, ..., y_K)^T$ is the true label of a pattern.

One may consider using another loss function. For example, it is known that in the case of imbalanced classes, the squared error loss function might work well [5], although in our case it did not work well.

Finally, in the study, the convolutional neural network was compared with the Support Vector Machine (SVM) algorithm [47]. SVM algorithm is a powerful general-purpose algorithm, based on a comparison of observed pattern to support vectors.

**Evaluation of detection algorithm**
To be able to judge the performance of the detection or classification algorithm, often sensitivity and specificity measures are used [6]. In the study, algorithm sensitivity (also called the true positive rate, the recall, or probability of detection), is defined as the probability of correctly anticipated seizure within a time horizon or epoch, and it is vital for judging the performance of seizure prediction. On the other hand, a classical definition of specificity (also called the true negative rate) in the seizure prediction context is the probability of correctly indicating a non-seizure state at any given time [7]. However, one can obtain very high sensitivity by simply tampering with a cut-off threshold, i.e. the point in the interval [0;1] when the judgment about the epoch changes from non-seizure to seizure is performed. Alternatively, another measure, for which the cut-off value is irrelevant, is a Receiver Operating Characteristic (ROC) curve [8] and a derived measure as an area under the ROC curve or Area Under Curve (AUC). ROC is not affected by the imbalance of classes (a very severe issue in all seizure detection cases), and it can be used to obtain an optimal cut-off value (e.g. optimality might be defined as a maximum of a sum of sensitivity and specificity).

In this study, ROC and AUC measures together with optimal sensitivity and specificity measures were used exclusively. AUC measure provided the information on how better the detection algorithm is as compared to random guessing of the classes or labels of epochs.

# Results and discussion

In this section, the results of the study and the main task of seizure detection and classification are discussed. First, the data sample is explored in terms of synchronisation (PLV), Shannon entropy and power spectrum features – discussion on the classification of classes is given, based on distributions of the features mentioned above. Next, the results of the trained classifier are presented. The main task was to classify each 1-second duration epoch to classes "seizure" and "non-seizure". The more specific results of the classification of all types of seizures defined in Table 1 are also presented and discussed.

Two additional subtasks of smaller scale are considered as well: 1) seizure detection using first 10 second of each seizure, i.e. annotations is modified in such a way, that first 10 second of the seizure epochs could be labeled as "seizure" and the rest as only "non-seizure" epochs; 2) seizure detection using 50 % of the testing set, when the training set is extended by the first 50 % of the test set, i.e. 50 % of EEG records for each patient in the test set was added to the training set. The classifier trained in the second subtask is patient-specific and is opposed to the patient-non-specific classifiers trained during the first subtask and other initial cases.

These 10 second and 50 % values are arbitrary and should not be taken as some kind of a rule. The additional options were chosen only for the sake of demonstration and evaluation of the classifier from various angles. The first subtask serves as an indication of how well the classifier picks up initial phases of seizures, while the second subtask has the purpose of evaluating our approach and CNN architecture to obtain patient-specific classifier, as opposed to the patient-non-specific classifier. Such kind of early detection or more accurate patient-specific detection, for example, is critical for successful intervention with a responsive neuro-stimulation device.

In each case, analysis of sensitivity to the input (not just AUC or sensitivity in relation to prediction) is presented as well. Sensitivity to the network inputs is defined as a squared gradient of a loss function, averaged over all testing samples. This sensitivity to the network inputs must not be confused with sensitivity as a measure of the classifier – which one is used should be clear from the context where it is used.

### Data analysis

In what follows, an overview of seizure detection task using the TUH EEG Corpus dataset, in terms of previously defined features of brain state, is given. This reflects the reasoning of classification to "seizure" vs "non-seizure" patterns.

Even though distributions cannot be a rule of thumb for detection, it is important to look at the distributions of various brain state features during the seizure and non-seizure periods. The distribution of each feature (Figure 2), raises concerns, that features are not capable of efficient differentiation between seizure and non-seizure classes. The phase locking feature (PLV), as a feature related to synchronisation and initially assumed highly useful for prediction, actually, poorly reflects different classes – its distribution only has a slight shift and increase in values for seizure class. The same or even worse is for the entropy, which is also a feature of synchronisation. Just energy (or power spectrum) feature carries more information to distinguish the two classes, although it is still not very well expressed.

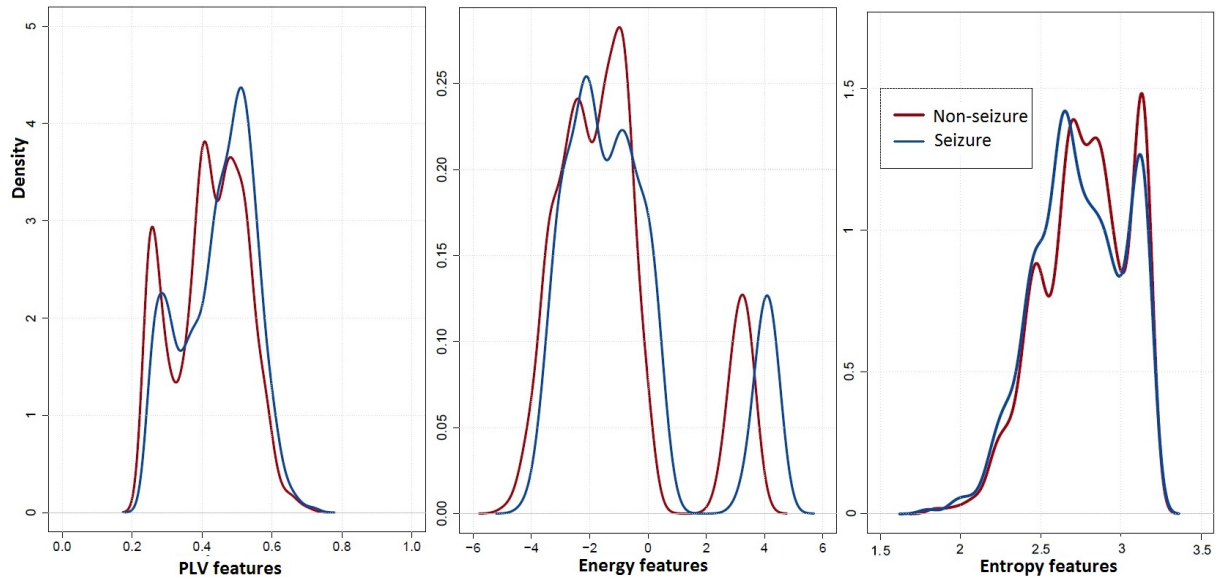

*Figure 2* Comparison of distributions of all features averaged over all dataset and separated between two classes. This figure compares three features in terms of probability distributions differentiated into two classes: the non-seizure and seizure. The features appear to be not very informative with regard to two classes (non-seizure and seizure). This is contrary to the observations in literature – many papers have shown that these features can differentiate well between seizure and non-seizure. The most likely reason that in this particular case these features do not show strong differentiation is that the EEG data used to construct them is very noisy, obtained from the highly variable population (age, health condition, gender variability), while other papers used data obtained under strictly controlled laboratory conditions.

Separation into different frequency bands (Figure 3) again shows minor differences in feature values. Thus, it is difficult to hope that the trained classifier will provide high accuracy, given that two classes are not very well distinguished using visual inspection of brain activity patterns. On the other hand, deep learning methods are known to have the property of learning features (deriving new features), which have high predictive power even though the interpretation of those new features cannot be interpreted like initially defined PLV, energy or entropy features.

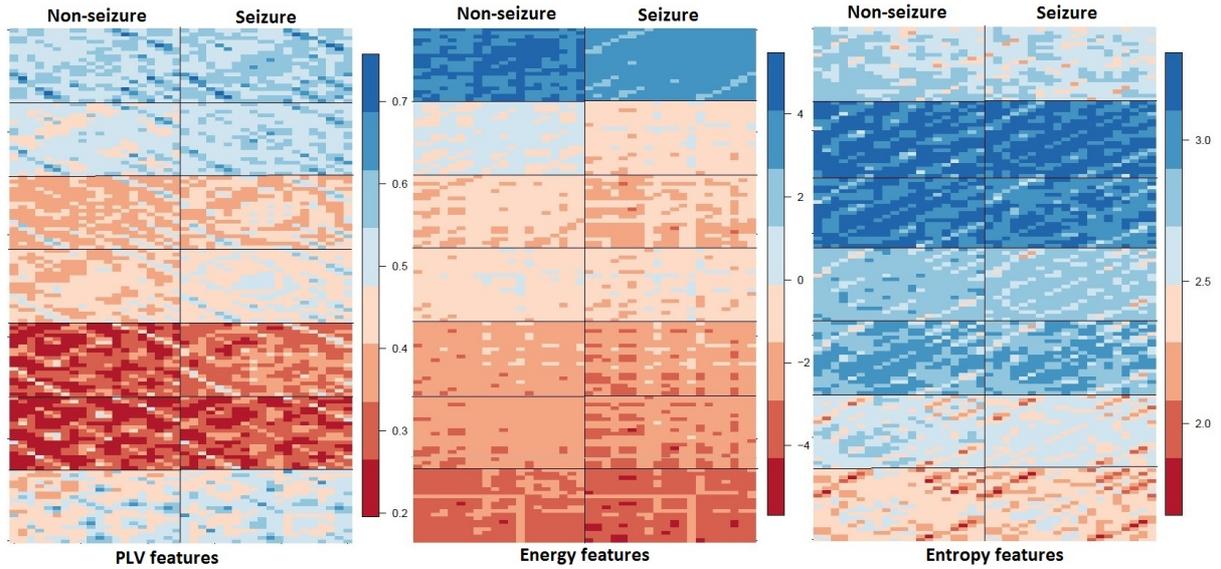

Figure 3 Comparison of all feature patterns, formed as described in "Patterns in brain activity" subsection, averaged over an entire training dataset and separated between two classes. Images from left to right: PLV, energy, and entropy averaged features. From top to bottom - there are 7 zones corresponding to 7 bandpass filtrations of EEG: [1-4] Hz, [4-7] Hz, [7-13] Hz, [13-15] Hz, [15-30] Hz, [30-45] Hz, [45-70] Hz. Each image has two sides – non-seizure and seizure. One can observe that different frequency bands differ for non-seizure and for seizure zones.

**Classification and detection results**

As the main task is the classification of seizure vs non-seizure epochs, initially, the feature patterns, formed and described previously, were fed to the convolutional neural network trained by the Adam Optimizer [4] with learning rate equal to 1E-3. However, the high imbalance of the classes (6 % of the time is in seizure state) caused great difficulty for the algorithm. This is so because distributions of seizure and non-seizure features are very similar (Figure 2) and high imbalance of the classes prevented proper learning. For example, there are many episodes with high synchronisation of the brain not only during the time interval with a seizure but also during a seizure-free time interval, which is much longer. Hence, the algorithm learns that higher phase locking value is not an indicator of seizure epoch. This effect is hardly reduced by the usual technique based on oversampling of the seizure epoch. Practically, it is simply not enough seizure epochs as compared to the non-seizure epochs. However, random under-sampling of the non-seizure epochs and such balancing of classes resulted in better detection accuracy.

Besides, 10-fold cross-validation is used to choose the architecture of the convolutional neural network. Cross-validation is done patient-wise, i.e. patients were divided into ten random groups. In each fold, a check is performed whether the accuracy is affected by the above-mentioned random under-sampling of non-seizure labels or not. It was observed, that there is no impact of the random selection of non-seizure labels on the results of classification. In addition, random under-sampling was done to have a 50 % / 50 % distribution of class labels. Several architectures were considered:

1. CNN1 architecture: one convolutional layer (10 filters of size 5x5), followed by max-pooling by factor of 2 followed by densely connected layer with 1000 neurons;
2. CNN2 architecture: two convolutional layers (10 filters of size 5x5), followed by max-pooling by factor of 2 followed by densely connected layer with 1000 neurons;
3. CNN3 architecture: two convolutional layers (10 filters of size 5x5) each followed by max-pooling by factor of 2 (e.g. first convolutional layer + max pooling + second convolutional layer + max pooling) followed by densely connected layer with 1000 neurons;
4. CNN4 architecture: two convolutional layers (10 filters of size 5x5) each followed by max-pooling by factor of 2 (e.g. first two convolutional layers + max pooling + third and fourth convolutional layers + max pooling) followed by densely connected layer with 1000 neurons;

The average AUC of 10-fold cross validation procedure are reported in Table 3. In addition, results for support vector machine are presented as well. The cross-validation was performed on the training dataset, by dividing it randomly into 10 equal parts, then the average of 10 AUC values is calculated.

*Table 3 10-fold crossvalidation accuracies expressed as AUC values for considered CNN classifiers and SVM.*

| Classifier | PLV | Energy | Entropy | PLV+ Energy | PLV+ Entropy | Energy+ Entropy | All |
|---|---|---|---|---|---|---|---|
| **CNN1** | 0.65 | 0.64 | 0.64 | 0.65 | 0.62 | 0.67 | 0.70 |
| **CNN2** | 0.67 | 0.66 | 0.65 | 0.71 | 0.67 | 0.70 | 0.74 |
| **CNN3** | 0.66 | 0.68 | 0.63 | 0.69 | 0.70 | 0.71 | 0.73 |
| **CNN4** | 0.68 | 0.65 | 0.68 | 0.70 | 0.68 | 0.69 | 0.72 |
| **SVM** | 0.62 | 0.51 | 0.59 | 0.63 | 0.63 | 0.60 | 0.64 |

Further follows more detailed analysis of testing results with best CNN architecture (CNN2). Although CNN2 and CNN3 architectures shows almost identical results, the simpler architecture (CNN2) was selected for further analysis.

For comparison, at first, each feature can be considered separately. This allows assessment of the explanatory power of each feature. Convolutional neural network applied to the phase locking value (PLV) feature patterns produced AUC = 0.67 value with optimal sensitivity and specificity being 0.62 and 0.64 respectively. Average energy features produced accuracy close to the PLV features' case (AUC = 0.66 with optimal sensitivity and specificity being 0.62 and 0.61). The similar results are for entropy features: AUC=0.65, with sensitivity and specificity being 0.59 and 0.62.

In addition to seizure detection, the authors checked how CNN detects eight different types of seizures. The results for each feature and each type of seizures are presented in Table 3. Combinations of features improve overall results only slightly: for PLV + Energy AUC = 0.71, for PLV + Entropy AUC = 0.67, for Energy + Entropy AUC = 0.7, for all features, i.e. PLV + Energy + Entropy AUC = 0.74. The sensitivity and specificity of the classification algorithm, trained on all features is 0.68 and 0.67 respectively.

It is particularly interesting to compare classifiers for different types of seizures. For each type of seizures, the evaluation was done by considering patterns of other types of seizures and non-seizure patterns as belonging to one class.

*Table 4 Detection results for each seizure type annotated in the evaluation dataset. Three measures (top to bottom) – AUC, sensitivity, and specificity values were reported.*

| Features | Focal non-specific seizure (0.6 %) | General non-specific seizure (16 %) | Simple partial seizure (45.75 %) | Complex partial seizure (26.18 %) | Absence seizure (0.68 %) | Tonic seizure (6.82 %) | Clonic seizure (1.5 %) | Tonic-Clonic seizure (3.93 %) |
|---|---|---|---|---|---|---|---|---|
| **PLV** (AUC=0.67) | **0.78** 0.72 0.74 | **0.74** 0.68 0.69 | **0.62** 0.6 0.58 | **0.67** 0.62 0.62 | **0.68** 0.67 0.62 | **0.75** 0.71 0.66 | **0.64** 0.57 0.67 | **0.39** 0.35 0.50 |
| **Energy** (AUC=0.66) | **0.88** 0.61 0.80 | **0.71** 0.65 0.69 | **0.62** 0.58 0.61 | **0.6** 0.61 0.5 | **0.64** 0.78 0.54 | **0.68** 0.69 0.58 | **0.54** 0.57 0.51 | **0.84** 0.76 0.77 |
| **Entropy** (AUC=0.65) | **0.86** 0.74 0.85 | **0.72** 0.68 0.65 | **0.65** 0.6 0.61 | **0.63** 0.58 0.59 | **0.63** 0.59 0.66 | **0.6** 0.64 0.5 | **0.46** 0.39 0.56 | **0.46** 0.58 0.38 |
| PLV + Energy (AUC=0.71) | 0.98 | 0.63 | 0.65 | 0.8 | 0.65 | 0.96 | 0.6 | 0.70 |
| | 0.95 | 0.65 | 0.64 | 0.74 | 0.76 | 0.89 | 0.52 | 0.56 |
| | 0.94 | 0.50 | 0.61 | 0.56 | 0.56 | 0.91 | 0.67 | 0.77 |
| PLV + Entropy (AUC=0.67) | 0.87 | 0.78 | 0.63 | 0.67 | 0.72 | 0.80 | 0.40 | 0.54 |
| | 0.76 | 0.70 | 0.57 | 0.63 | 0.70 | 0.73 | 0.30 | 0.65 |
| | 0.82 | 0.71 | 0.62 | 0.60 | 0.62 | 0.70 | 0.60 | 0.42 |
| Energy + Entropy (AUC=0,70) | 0.86 | 0.72 | 0.65 | 0.63 | 0.63 | 0.6 | 0.46 | 0.46 |
| | 0.74 | 0.68 | 0.6 | 0.58 | 0.59 | 0.64 | 0.39 | 0.58 |
| | 0.85 | 0.65 | 0.61 | 0.59 | 0.66 | 0.5 | 0.56 | 0.38 |
| PLV + Energy + Entropy (AUC=0.74) | 0.97 | 0.63 | 0.73 | 0.78 | 0.62 | 0.95 | 0.55 | 0.70 |
| | 0.90 | 0.62 | 0.69 | 0.72 | 0.80 | 0.89 | 0.50 | 0.65 |
| | 0.92 | 0.58 | 0.66 | 0.72 | 0.50 | 0.90 | 0.60 | 0.60 |

Detection accuracy for simple partial seizures (the most frequent type of seizures) is homogeneous between different features and combinations of them. A more significant improvement can be achieved by combining all features although the improvement is counteracted by significantly increased dimensionality and the computational burden.

Another frequent type is a complex partial seizure, and here no improvement is made with learning from all features. In this case, the feature of synchronisation (i.e. PLV) and energy over different bands of frequencies is sufficient to achieve AUC=0.8.

Also, the high accuracy of results for non-specific focal seizures probably is due to the ability of the classifier to distinguish its patterns accurately. Significantly poorer results for other rare types (absence and clonic seizures) allows rejecting the hypothesis that high results for non-specific focal seizures are solely due to the small sample effect; surely small sample plays its role, but probably not definitive – the deeper investigation is out of the scope of this paper. Further analysis of the table reveals that energy is a highly informative feature for this type of seizures, while entropy provides almost no additional information.

The sensitivity of loss function (here defined as the square of the loss function gradient taken w.r.t. input features) to various features provides an additional level of information. Results show (Figure 4), that phase locking value carries most information about seizure/non-seizure classes in the middle and higher frequencies (7-70 Hz), while entropy (also a feature of phase synchronisation) is more sensitive to the lower frequencies in the range of 1-30 Hz. On the other hand, energy considering very low frequencies (1-4 Hz) is almost irrelevant with most of the information contained in the middle frequency ranges (4-30 Hz). Also, each feature has a small group of channels (specific zones) mostly involved in the discrimination between seizure and non-seizure classes – groups of channels are different for different features. This provides implications to the dimensionality reduction – one could use a smaller number of frequency ranges as well as channels. This would be of great importance for the seizure detection and prediction tool, as such reduction could potentially reduce the calculation time and thus decrease the prediction timing.

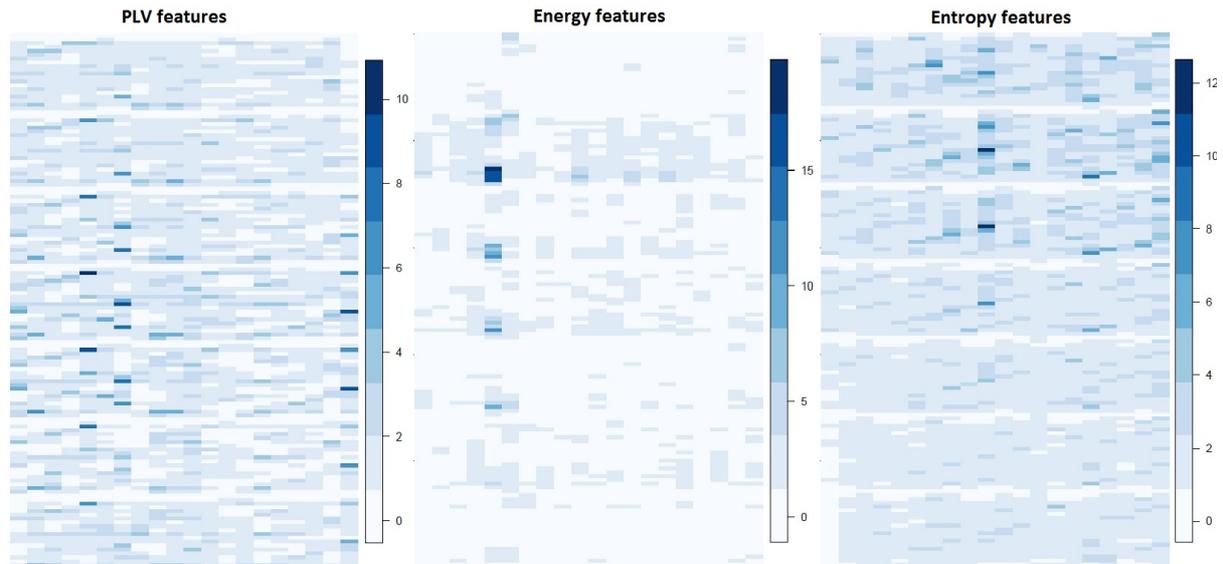

Figure 4 Loss function sensitivity to different features. The structure of the images is the same as in figure 2, but here average gradients of loss function are taken w.r.t. each entry of the brain activity pattern. The higher the value, the higher the sensitivity. High sensitivity indicates that the classifier strongly reacts to the change in that particular feature entry and therefore those entries are responsible for the detection of seizures. Low sensitivity indicates that no matter the change of that particular feature entry, the classifier will no react strongly to that change. This indicates that those feature entries carry only a small amount of information. Given such information, it might be useful to only consider entries with the highest sensitivity and in this way to reduce the dimensions boosting the speed of calculations.

Further, authors tested how difficult it is to detect the start of the seizure, i.e. the seizure during the first 10 seconds. Detection results were even more pessimistic (Figure 5). The highest AUCs were achieved combining energy and entropy features, besides PLV features gave no further boost in the detection accuracy. Synchronisation related PLV feature, alone and in combination with entropy, poorly detected the start of the seizure. This might be due to the desynchronization during the initial phase of the seizure – this kind of feature is of value for the detection of later phases of seizure. However, one of the most important issues in epilepsy is the prediction of seizure onset, while the later phases are less relevant. Hence, application of energy features should be a priority in seizure detection. Detection of seizure during the first 10 seconds just before the seizure onset (as judged by a trained clinician) failed completely: AUC values were even lower and concentrated around 0.5.

From the above results, one could make a general conclusion: seizures are not easily predictable with satisfactory accuracy by current state-or-art classification methods, as used on clinical EEG recordings. That is probably an answer to the observation, that seizure detection alarm tools used in hospitals have a very high false positive rate, as opposed to the high accuracy of classifiers, reported in many research papers, but trained and tested on EEG had been obtained under non-clinical conditions. Then a valid question would be: is it at all possible to have such an algorithm, which detects various types of seizures for patients of different ages and suffering different illnesses? It is challenging and at least hardly possible with the current widespread approach when many different features are constructed from EEG simply dividing recordings into short time windows and without the use of knowledge about the dynamics of seizures and relations between different phases and features.

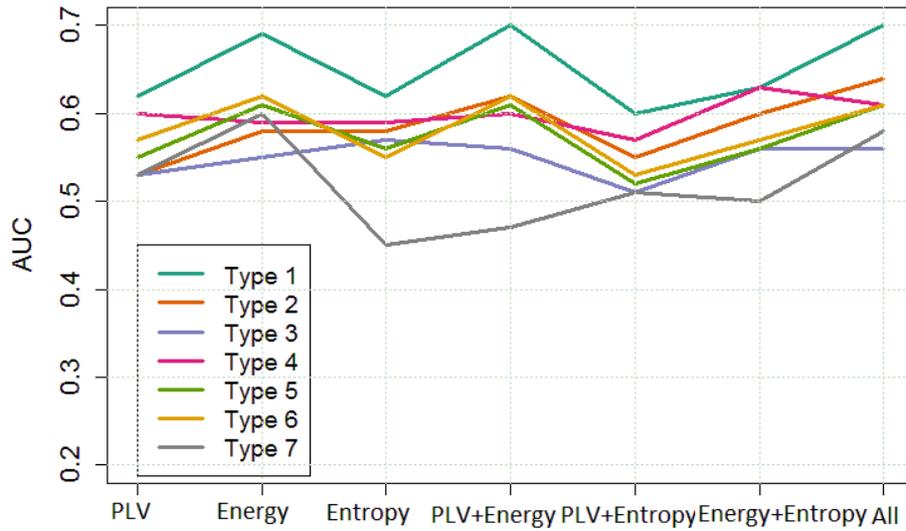

Figure 5 AUC results for each type of seizure (see table 1 for seizure type names). From this figure, one can observe that accuracies of detection are very similar for all except Type 1 and Type 7 seizures. We discarded the 8[th] type of seizures (tonic-clonic) from this figure because for every combination of features the detection accuracy (in terms of AUC) was less than 0.5 indicating that none of the features is able to differentiate this particular type of seizures.

On the other hand, patient-specific detection system could be based on a classifier related to the one investigated above. In this case, 50 % of each patient's EEG records (i.e. the first 50 % of timing interval for each patient) from test dataset is added to the training set. Then the convolutional neural network is again trained to give much more optimistic results (Figure 6). In this case, AUC is above 0.85 for various features and combinations (even achieving AUC=0.9 using all types and all features). When each type of seizures is considered separately, a large deviation is observed, especially for Clonic seizures (i.e. type 7), while accuracy for detection of other types of seizures had similarly high AUC values.

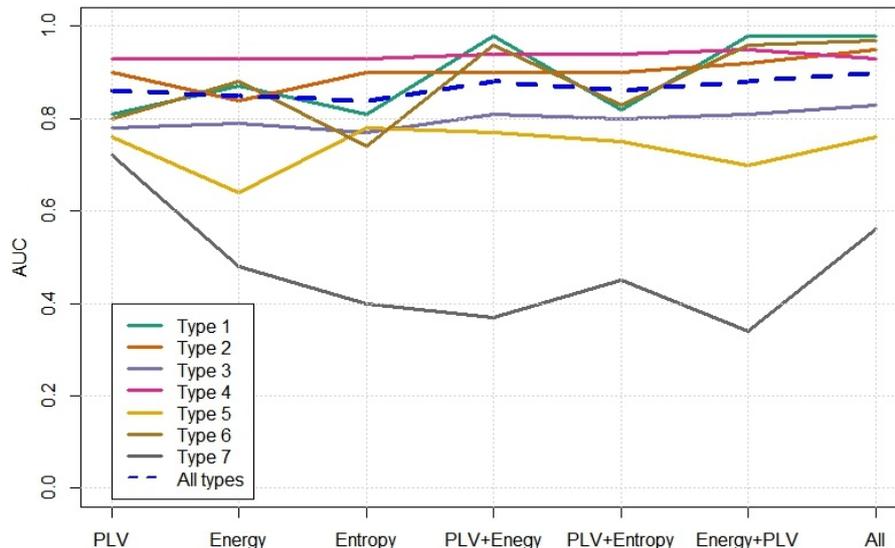

Figure 6 Results of patient-specific classifier: The first 50 % of EEG records in test dataset for each patient was used to train the classifier (in addition to the previously used training dataset). One can observe that even though for the first six types of seizures detection improved, for Type 7 the effect was the reverse. We discarded the 8[th] type of seizures (tonic-clonic) from this figure because for every combination of features the detection accuracy (in terms of AUC) was less than 0.5 indicating that none of the features is able to differentiate this particular type of seizures.

# Summary and Conclusions

In this paper, the clinical data from Temple University Hospital EEG Corpus database is studied. The main task of seizure detection and classification is solved by the application of convolutional neural network – state-of-the-art method, which has demonstrated to be of superior accuracy among other previously applied methods. Images of brain activity were constructed using different features, i.e. phase locking value, entropy, energy. These images then enabled the application of various deep learning methods and specifically focused on CNN.

The results are rather pessimistic, and we conclude that for clinical data one cannot construct a universal seizure detector in a way how patient-specific data had been used for the past decades (i.e. extracting various features, e.g. energy and synchronisation, and then simply glueing them together). One could even go further and conclude that an accurate tool for clinical/hospital usage will not be constructed unless the shift of paradigm occurs and more attention will be paid to this challenge.

In our case for such universal detector, the sensitivity and specificity reached 0.68 and 0.67, accordingly, and Area Under the Curve was around 0.74 (with slight variations depending on the type of seizures and features used). When different types of seizures are considered separately, a very large variation of results was identified: from very poor AUC=0.39 for tonic-clonic seizures (with PLV features) up to very high AUC=0.98 for non-specific focal seizures (with PLV and Energy features).

In future investigations, dimensionality reduction based on cost function sensitivity to inputs could increase the AUC values. However, the increase might be only marginal. The above-reported results for the universal seizure detector imply a certain lack of detectability of seizures, if it is based on the energy and synchronization features alone.

On the other hand, the patient-specific classifier (i.e. a classifier trained on each patient data) could be useful in personal or even mobile applications. For personal data treatment cases, AUC is around 0.9, which could be even slightly increased by dimensionality reduction.

# Compliance with Ethical Standards

**Conflict of interest** The authors have no conflict of interest.
**Ethical Approval** This article does not contain any studies with human participants performed by any of the authors.
**Funding** Tomas Iesmantas was supported by the postdoctoral fellowship grant (2016-2018) from Kaunas University of Technology and Department of Mathematics and Natural Sciences.

**Tomas Iešmantas** completed his Ph. D. in technological sciences (Bayesian assessment of reliability and vulnerability of complex systems, 2016, Lithuanian energy institute). Currently he is working at the Kaunas University of Technology (Department of Mathematics and Natural Sciences). Recent interests are in deep learning and applications in medicine. More specifically, for biosignals as well medical images – epilepsy, cardio signal classification, MRI X-ray image classification and segmentation, etc. He is currently participating in several research projects and collaborating with other academic institutions, industries and governmental agencies.

**Robertas Alzbutas** completed his Ph. D. in technological sciences (Risk minimization and reliability control under uncertainty, 2004, Lithuanian energy institute). Currently he is working at the Kaunas University of Technology (Department of Mathematics and Natural Sciences). Recent interests are in connected health and precision medicine areas. His research include application of mathematical models in wearables for health state monitoring.